\documentstyle[psfig]{l-aa}
\def\ea{et~al.\ }
\def\src{SAX~J1808.4-3658}

\begin{document}
\thesaurus{01(08.09.2 \src; 08.14.1; 13.25.1)}

\title{Discovery of the X-ray transient \src, \\a likely low-mass X-ray binary }
\author{J.J.M.~in~'t Zand\inst{1} \and J. Heise\inst{1}
 \and J.M.~Muller\inst{1,2} \and A. Bazzano\inst{3}
 \and M. Cocchi\inst{3} \and L. Natalucci\inst{3} \and P. Ubertini\inst{3}
 }
\offprints{J.J.M.~in~'t Zand}

\institute{   Space Research Organization Netherlands, Sorbonnelaan 2,
              3584 TA Utrecht, the Netherlands
         \and
                BeppoSAX Science Data Center, Nuova Telespazio,
                Via Corcolle 19, 00131 Roma, Italy
         \and
                Istituto di Astrofisica Spaziale (CNR), Area Ricerca Roma Tor
		Vergata, Via del Fosso del Cavaliere, 00133 Roma, Italy
                        }
\date{Received, accepted }
\maketitle

\begin{abstract}

We report the discovery of a fairly bright transient during observations
with the Wide Field Cameras on board the BeppoSAX satellite in September 
1996. It was detected at a peak intensity of 0.1 Crab (2 to 10 keV) and
lasted between 6 and 40 days above a detection threshold of 2~mCrab. 
Two very bright type I X-ray bursts were detected from 
this transient in the same observations.
These almost certainly identify 
this X-ray transient as a low-mass X-ray binary with a neutron star as
compact component. The double-peaked time history of both bursts at high 
energies suggests a peak luminosity close to the Eddington limit. Assuming
this to be true implies a distance to this object of 4~kpc. 
\end{abstract}

\section{Introduction}
\label{secintro}

Currently a program is carried out to regularly monitor the galactic bulge in 2 to
25 keV X-rays with the Wide Field Cameras (WFC) on board the
BeppoSAX satellite. The main purpose of this program is to monitor 
relatively weak and short transient activity from various types of sources,
in particular low-mass X-ray binaries (LMXBs). 
According to a recent count (Van Paradijs 1995), near to 30\% of all $\sim$130
known LMXBs are transient in nature. This population is concentrated in
the sky towards the direction of the galactic center (e.g., 
Van Paradijs \& White 1995, White \& Van Paradijs 1996). 

We here present the discovery of a relatively bright transient during
observations in September 1996, which exhibited X-ray bursts,
and discuss its timing and spectral behavior in X-rays.
In Sect.~\ref{secobs}, we discuss the observations, in Sect.~\ref{sectdetection}
the detection and position of the
transient, in Sect.~\ref{secslow} trends in the intensity
and spectrum, in Sect.~\ref{secbursts} the two bright X-ray bursts
that were detected and in Sect.~\ref{secdisc} we evaluate the data.

\section{Observations}
\label{secobs}

\begin{table}[b]
\caption[]{Main characteristics per BeppoSAX-WFC camera}
\begin{tabular}{ll}
\hline
Field of view 		&  40$^{\rm o}$ $\times$ 40$^{\rm o}$ full width to zero response\\
			& (3.7\% of entire sky)\\
Angular resolution      &  5 arcmin\\
Source loc. accuracy&  $>$0.6 arcmin (68\% conf. level)\\
Detector technology	&  Multi-wire prop. Xenon counter\\
Photon energy range	&  2 to 25 keV\\
Energy resolution	&  18\% at 6~keV\\
Time resolution		&  0.5~ms\\
\hline
\end{tabular}
\label{wfcpars}
\end{table}

The WFC instrument (Jager et al. 1997) comprises 2 identically designed coded 
aperture cameras on the BeppoSAX satellite (Boella \ea 1997) which was
launched in April 1996. The main characteristics per camera are presented in 
Table~\ref{wfcpars}. The field of view (FOV)
of this instrument is the largest of any flown X-ray imaging device.
The moderate angular resolution of 5 arcmin (full-width at half maximum)
does not pose severe problems with
regards to source confusion for the sensitivity of this instrument, even in
the galactic bulge field. There are only few X-ray sources persistently 
brighter than
10 mCrabs (2 to 10 keV) or $\sim2~10^{-10}$~erg~s$^{-1}$~cm$^{-2}$ which are
closer than 5 arcmin to the nearest such neighbor.

In the context of sensitivity it is important to note that an imaging
device based on the coded aperture principle has one very basic difference
with direct-imaging devices such as X-ray mirror telescopes: there is cross-talk
between FOV positions that are well beyond one angular resolution distance.
For the WFC, this degrades the sensitivity to any sky position within
20$^{\rm o}$ from a bright X-ray source. This
has a relatively large impact on observations of the crowded galactic bulge
field where the sensitivity is about two times less than at high galactic
latitudes and far from bright sources.

The point source sky position accuracy of the instrument is an order of
magnitude better than the angular resolution. It is limited by systematic
errors. In-flight calibration (In 't Zand et al. 1997) 
determined that the systematics are near-Gaussian distributed with a 68\% 
confidence error circle radius of 0.6~arcmin.

The galactic bulge was, during 1996, targeted for 22 days in 
August, September and October. The total exposure time of these observations 
is about 7$\times10^5$~s. The source reported in this paper is 9.1~degrees from 
the galactic center, well within the field of view of BeppoSAX-WFC. Therefore, 
the coverage of the source is near to complete within this data set.

\begin{figure}[t]
\psfig{figure=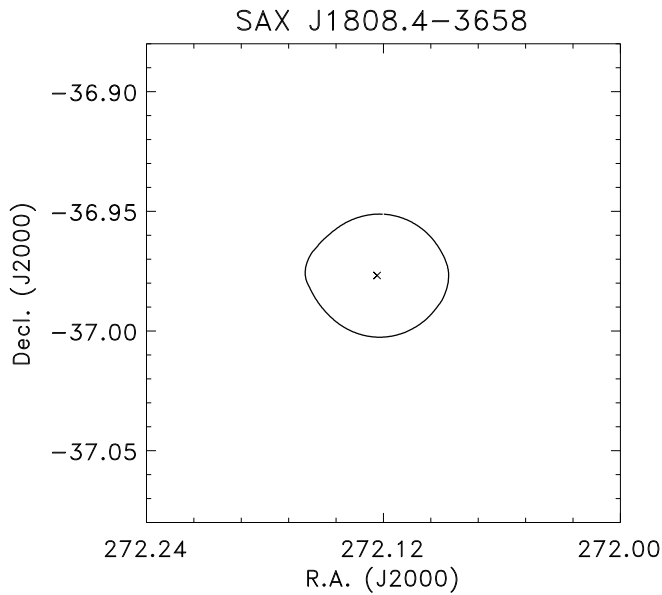,width=\columnwidth,clip=t}

\caption[]{Error box of steady emission (99\% confidence level). The cross
indicates the best fit position
\label{errorbox}}
\end{figure}

\begin{figure}[t]
\psfig{figure=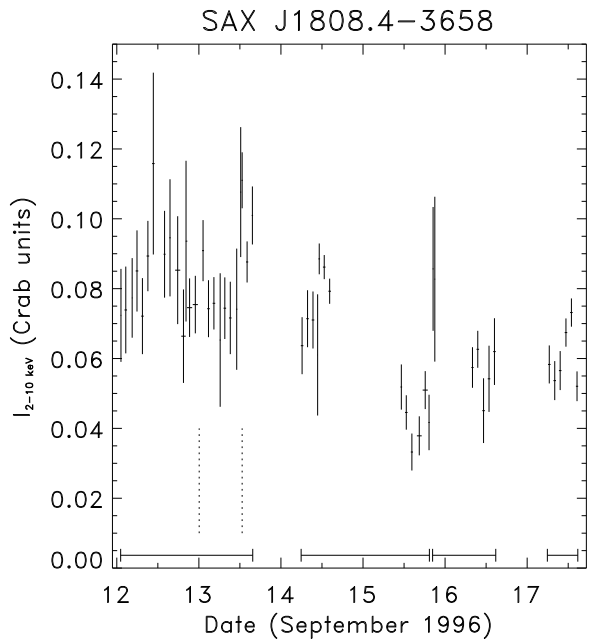,width=\columnwidth,clip=t}

\caption[]{Light curve of \src between 2 and 10 keV. The horizontal bars 
indicate the time intervals used in the spectral modeling (see 
Sect.~\ref{secslow}), the vertical dashed lines indicate the times
of two X-ray bursts (see Sect.~\ref{secbursts})
\label{1808slowlc}}
\end{figure}

\begin{figure}[t]
\psfig{figure=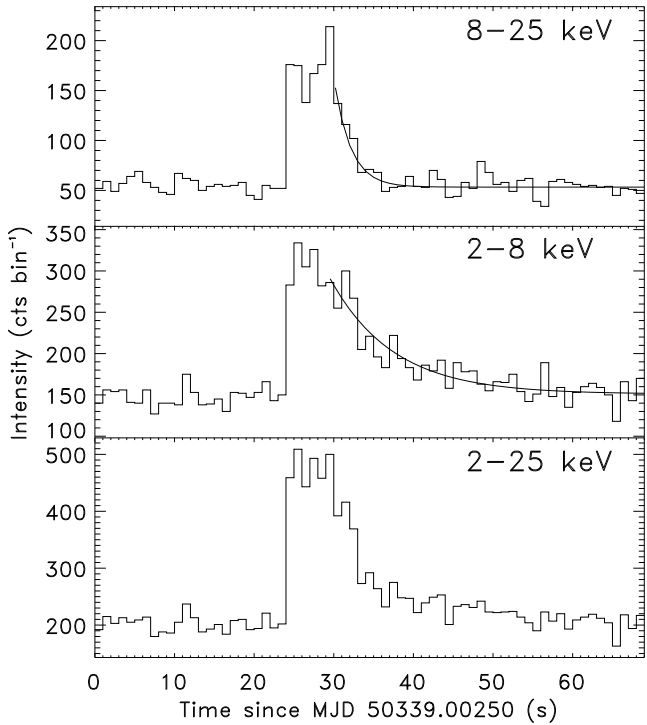,width=\columnwidth,clip=t}

\caption[]{Time profile of the first burst, per each of two bandpasses and for
the complete WFC bandpass. The bin time is 1~s. The smooth
curves are exponential models for the appropriate time profiles (see
text)
\label{burst1}}
\end{figure}

\begin{figure}[t]
\psfig{figure=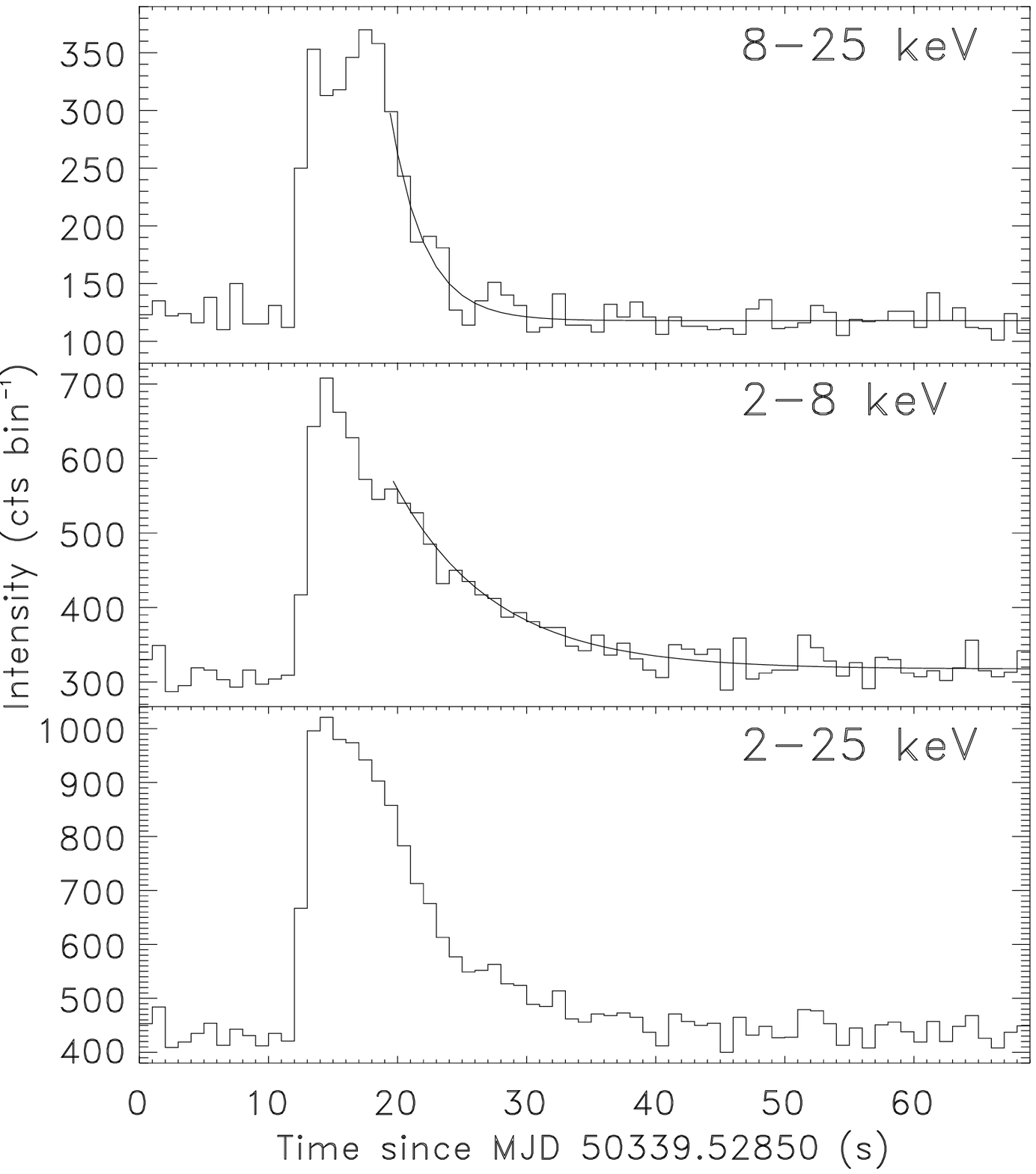,width=\columnwidth,clip=t}

\caption[]{Time profile of second burst from \src
\label{burst2}}
\end{figure}

\section{Detection and position}
\label{sectdetection}

The transient was detected during the observations on September 12
through 17, 1996, for a total exposure time of 1.3~10$^5$~s.
Fig.~\ref{errorbox} shows the error box of the source. This was
determined from 0.15~d worth of data with high statistical quality. 
The best fit position is

\vspace{0.3cm}
\noindent
R.A.~=~18h~08m~29s, Decl.~=~-36$^{\rm o}$~58\farcm6 (J2000.0).

\vspace{0.3cm}
A check against X-ray catalogs in the Simbad database revealed no known 
X-ray source in this error box or, for that matter, within 5 arcmin from 
the best fit position. We designate the source as \src.

The source was not detected during the previous observation of the same sky 
field, 13 days before September 12. For the complete observation run between 
August 21 and 30, 1996, the 3$\sigma$ upper limit on the intensity is 1.5~mCrab 
(2 to 10 keV). Also, \src\ 
was not detected during the following observation between October 10 and 12, 
1996, for which the 3$\sigma$ upper limit is 2.0~mCrab. Therefore, we
conclude that the transient was active between 6 and 40 days.

\section{Light curve and spectrum}
\label{secslow}

\begin{table}[b]
\caption[]{Results of spectral modeling of steady emission}
\begin{tabular}{lrrr}
Model$^{\rm a}$ & Model parameter & $N_{\rm H}$ (10$^{22}$ cm$^{-2}$) & $\chi^2_{\rm r}$ \\
\hline
PL  & $\Gamma=2.17\pm0.04$ & $<1.1$   & 0.95 \\
BR & $kT=7.4\pm0.4$ keV  & $<0.9$ & 1.00 \\
\hline
PL  & $\Gamma=1.99\pm0.08$ & $<1.2$   & 0.88 \\
 &                  $2.06\pm0.07$ & \\
 &                  $2.36\pm0.16$ & \\
 &                  $2.35\pm0.09$ & \\
BR  & $kT=9.4\pm1.1$ keV & $<1.1$ & 0.90 \\
 &                  $8.5\pm0.9$ & \\
 &                  $6.2\pm1.0$ & \\
 &                  $6.0\pm0.6$ & \\
\hline
\end{tabular}

\noindent
$^{\rm a}$PL~=~power law spectrum 
(i.e., flux in phot~s$^{-1}$cm$^{-2}$keV$^{-1}$ is proportional to $E^{-\Gamma}$ 
where $E$ is the photon energy), 
BR~=~Thermal bremsstrahlung spectrum
\label{tabspectra}
\end{table}

In Fig.~\ref{1808slowlc} the light curve of the source is plotted in bins 
equal to one BeppoSAX orbit ($\sim103$~min) which is the highest time resolution 
the statistical quality of the data permits. The peak intensity is
equivalent to about 0.1~Crab between 2 and 10 keV, on a 1 day time scale
this is about 0.08 Crab. There appears to be a declining
trend by about a factor of 1.5 over 5.5~days. If this trend would follow an
exponential decay (like in many LMXB transients, see Chen, Shrader \& Livio
1997, but not verifiable for this
transient) the 1/e decay time would be 14~d. If the brightness of the 
transient would indeed have followed such an exponential decay we can
estimate what the maximum intensity could have been assuming the 
transient peaked immediately after the previous observation 
on August 30 when it was not detected: 0.2~Crab units in 2 to 10 keV.

Spectra were accumulated for each of 16 observation periods in 18 channels
between 3 and 10 keV (this restriction was imposed by the current status
of the spectral calibration of the instrument). The spectra
were fitted with two single-component models. Table~\ref{tabspectra} presents 
the results.
The first part of the table presents the results
leaving free per observation period only the normalization (267 degrees of
freedom), the second part does so leaving free the index parameter (power
law photon index or temperature k$T$) over 
four different time intervals (264 d.o.f.). The four intervals
are indicated in Fig.~\ref{1808slowlc}. In all cases the hydrogen column
density of cold interstellar matter $N_{\rm H}$ (applying the absorption model 
by Morrison \& McCammon 1983) is a single free
parameter and identical over all times. Three conclusions 
can be drawn from these results:
the spectral data does not favor either model, spectral softening 
is occurring over the 5.5~d of observation, and $N_{\rm H}$
may be as high as about $\sim10^{22}$~cm$^{-2}$.

\section{X-ray bursts}
\label{secbursts}

\begin{table}[b]
\caption[]{Characteristics of two bursts}
\begin{tabular}{lll}
\hline
Parameter & Burst 1 & Burst 2 \\
Start time (MJD)        & 50339.00278   & 50339.52864 \\
Instrument              & WFC2                  & WFC2          \\
Peak intensity (Crab, 2-8 keV) & 3.8 $\pm$ 0.2   &  3.5 $\pm$ 0.1 \\
Decay time $\tau$ (s, 2-8 keV)  & 7.8$\pm$1.2          & 7.5$\pm$0.8 \\
Decay time $\tau$ (s, 8-25 keV) & 2.1$\pm$0.3           & 2.6$\pm$0.3\\
BB color temperature 0-7 s (keV)  & 2.8$\pm0.2$ & 2.2 $\pm$ 0.1 \\
BB color temperature 7-14 s (keV) &  -- & 1.7 $\pm$ 0.08 \\
\hline
\end{tabular}
\label{tabbursts}
\end{table}

Two very bright bursts were detected from this source, the time histories in 2 
energy bands (2 to 8 and 8 to 25 keV) are given in Figs.~\ref{burst1} and
\ref{burst2}, and some characteristics are listed in Table~\ref{tabbursts}. 
The time histories are of all counts in the subsection of the detector 
illuminated by \src. 
Both bursts appear quite similar. They have identical durations and similar
peak intensities. Also, both bursts have the same kind of double-peak behavior
particularly at high energies.

Exponential functional forms have been fitted to the time histories of
both bursts in each of the two bandpasses and well after the double-peaked 
period, leaving free the peak intensity, decay time and background level. 
The results have been drawn in Figs.~\ref{burst1} and \ref{burst2}.
The 1/e decay times $\tau$ thus found (see Table~\ref{tabbursts}) clearly 
demonstrate a softening whose evolution is identical for both bursts. 

The spectrum of the first 7~s of both bursts could be satisfactorily fitted 
with a black body model spectrum with a color temperature as indicated in 
Table~\ref{tabbursts}. The same applies to the second interval of 7~s of
the second burst (the statistical quality of the data does not permit an
accurate spectral modeling of this interval for the first burst). A black 
body spectrum gives, when assuming isotropic emission,
a direct relationship between the average radius of the emitting sphere
$R_{\rm km}$ (in units of km) and the distance $d_{\rm 10~kpc}$ (in units
of 10~kpc). All three spectral fits are consistent with 
$R_{\rm km}/d_{\rm 10~kpc}=19\pm5$. This number still needs to be
corrected for gravitational redshift and for the conversion of color 
temperature to true black body temperature. These corrections, which 
counteract each other and are anticipated not to exceed a factor of 2, 
will be dealt with elsewhere.

\section{Discussion}
\label{secdisc}

Of the two types of X-ray bursts found in many X-ray binaries (see review by 
Lewin, Van Paradijs \& Taam 1995), 
type I bursts are attributed to thermonuclear flashes on or near a
neutron star surface. Detection of type I bursts is, therefore,
a strong indicator for a neutron star. One diagnostic clearly distinguishes
type I from the other type of bursts: only type I bursts exhibit spectral 
softening. A further characteristic of type I bursts is that they
have black body spectra with temperatures up to a few keV. 
Thus, we can identify the two bursts reported here as type I bursts
and conclude that there is strong evidence for the neutron star
nature of this X-ray source.

Type~I X-ray bursts have been seen from 42 galactic X-ray binaries 
according to Van Paradijs (1995). Although not all of them have 
confirmed optical counterparts, those 19 that do are all LMXBs. 
All 42 bursters have been classified as LMXBs, directly through the
identification of the optical counterpart or indirectly through 
characteristics of the X-ray emission or association with a globular 
cluster. It is, therefore, almost certain that \src\ too is a LMXB with an 
as yet unidentified optical counterpart.

The double-peaked nature of both bursts at high energies is
indicative of near-Eddington luminosities (e.g., Lewin~\ea 1995).
This is supported by the black body temperatures which are similar
to the (likewise high) values obtained for other bursts that reach the
Eddington limit (e.g., Lewin at al. 1995).
If interpreted as such, an estimate can be obtained of the distance.
Assuming a 1.4~M$_{\odot}$ neutron star with an Eddington limit of
2~10$^{38}$~erg~s$^{-1}$ and an observed peak bolometric flux of
$(1.3\pm0.3)~10^{-7}$~erg~s$^{-1}$cm$^{-2}$, the distance is 4~kpc.
The galactic latitude of -8.1$^{\rm o}$
makes a distance closer than the galactic center (8.5~kpc) indeed likely.
Such a close distance suggests a reasonable perspective to find an optical
counterpart and we urge optical observers to follow up on the position 
here published. The observed peak flux of the steady emission is about 
2~10$^{-9}$~erg~s$^{-1}$cm$^{-2}$ in 2 to 10 keV which for the
power law spectrum extrapolates to a 0.4 to 10~keV luminosity of
($6\pm2$)~10$^{36}$~erg~s$^{-1}$ and for the thermal bremsstrahlung 
spectrum to $(5\pm1)~10^{36}$~erg~s$^{-1}$. The uncertainty in these
numbers is due to that in $N_{\rm H}$, a distance of 4~kpc is assumed.
The 0.4 to 10 keV peak luminosity is somewhat low though not unheard of
within the group of LMXB transients (e.g., Chen~et~al. 1997). 
The distance of 4~kpc implies a burst emitting sphere radius of 8~km. 
This supports the neutron star identification.

It is interesting to note that \src\ was not initially reported from 
data obtained with the all-sky monitor on board the Rossi X-ray Timing Explorer. 
The source was probably too near to
the sensitivity of this instrument which for sources in uncrowded fields
is about 50~mCrabs per dwell (3$\sigma$, Levine \ea 1996). The closest
bright source is X1822-371 at 3.5~degrees. Therefore, it appears that
the source was never much brighter than the peak intensity
observed with BeppoSAX-WFC.

\begin{acknowledgements}
We thank A. Klumper, M. Savenije, J. Schuurmans and G. Wiersma at SRON for 
software support during the analysis, and the staff of the 
BeppoSAX {\em Satellite Operation Center} and {\em Science Data Center} 
for the help in carrying out and processing
the Galactic Center observations with the WFC. The BeppoSAX satellite is 
a joint Italian and Dutch program. A.B., M.C., L.N. and P.U. thank Agenzia 
Spaziale Nazionale ASI for support. This research has made use of the Simbad 
database, operated at CDS, Strasbourg, France.
\end{acknowledgements}

\vspace{0.35cm}
\noindent
{\it Note added in proof.} 
A recent analysis of the ASM data using the position of the WFC detection 
reported here has revealed a detection (Remillard 1997, private communication) 
and confirmed our suspicion that the true peak intensity could not have been 
much higher than observed with WFC. In fact it was comparable. The transient 
rose above the ASM detection level for cataloged sources (about 20~mCrab in 2 
to 12 keV for one day of observations) on about September 8, 1996, and 
remained detectable for about 20 days. 

\end{document}